\documentclass[a4paper,11pt]{article}
\pdfoutput=1 
\usepackage{jcappub} 
\usepackage[T1]{fontenc} 
\title{\boldmath $\gamma$ rays run on time}

\author{Daniel Beltr\'an Mart\'{\i}nez,}
\author{Felipe J. Llanes-Estrada$^1$\note{Corresponding author.} and\\ }
\author{Gloria Tejedor-Garc\'{\i}a}

\affiliation{Departamento de F\'{\i}sica Te\'orica e IPARCOS, Univ. Complutense de Madrid.\\
Plaza de las Ciencias 1, Facultad de Ciencias F\'{\i}sicas, 28040 Madrid, Spain.}

\emailAdd{fllanes@ucm.es}

\abstract{Significant absorption of radiation is usually accompanied by refraction.
This is not the case for $\gamma$ rays travelling cosmic distances. We show that the real
and imaginary parts of the refraction index are indeed commensurable, as they are related
by dispersion relations, but when turning to physical observables, the (finite) optical depth is way larger than the (infinitesimal) time delay
of the gamma rays relative to gravitational radiation. \\
The numerically large factor solving the apparent contradiction is 
$E_\gamma/H_0$ arising from basic wave properties (Bouguer-Beer-Lambert law) 
and the standard cosmological model, respectively. }

\begin{document}
\maketitle
\flushbottom

\section{Motivation}
\label{sec:intro}

The gravitational wave signal of GW170817~\cite{LIGOScientific:2017vwq} was followed 1.7 seconds later by the 
$\gamma$ ray burst GRB170817A~\cite{LIGOScientific:2017ync,LIGOScientific:2017zic}. This is an awe-inspiring result, as the 
$\gamma$ rays were only delayed less than two seconds in a 130-million light-year travel from the
host galaxy NGC4993. Such small delay is adscribed to the source geometry~\cite{Salafia:2017hfr}, 
where the emission of nuclear matter is for an instant opaque; when it releases radiation,
an observer that is not perfectly on axis will perceive a small extra path.
For the rest, it is believed that the binary neutron star event must have been near the host galaxy's
edge so that there was  little matter in the way (that would cause Compton scattering). 

This precise timing of
gravitational waves and $\gamma$ radiation entails that they have precisely the same speed to a part
in $10^{-15}$ and has lead to interesting tests of General Relativity.

Multimessenger astronomy combining, among others, gravitational waves and gamma radiation will continue
at the fore when third generation gravitational wave detectors are constructed.
The Einstein Telescope~\cite{Punturo:2010zz}, particularly, is expected to reach large $z$ events, even of order 10. In this article
we will find that $\gamma$s are not measurably delayed respect to GWs at even the largest $z$. A different
question is their absorption: $\gamma$s of energy around 100 GeV and above are strongly absorbed, to the point
that a ``$\gamma$ horizon'' appears~\cite{Saldana-Lopez:2020qzx}. For a 100 GeV photon this is at about $z\simeq 1.2-1.4$; for a 50 GeV photon, rather $z\sim 3$. At this horizon the optical depth $\tau$ becomes of order 1, so that the photon count is suprressed by $\frac{1}{e}$ as per
\begin{equation}
\left(\frac{dN}{dE} \right)_{\rm observed}=\left(\frac{dN}{dE} \right)_{\rm emitted} \times e^{- 
 \tau(E,z)}\ .
\end{equation}
(Empirical studies to extract it from blazar data require an additional correction constant $b$ in the exponent that is of no concern in this work.)

\begin{figure} 
\begin{center}
\includegraphics[width=0.7\textwidth]{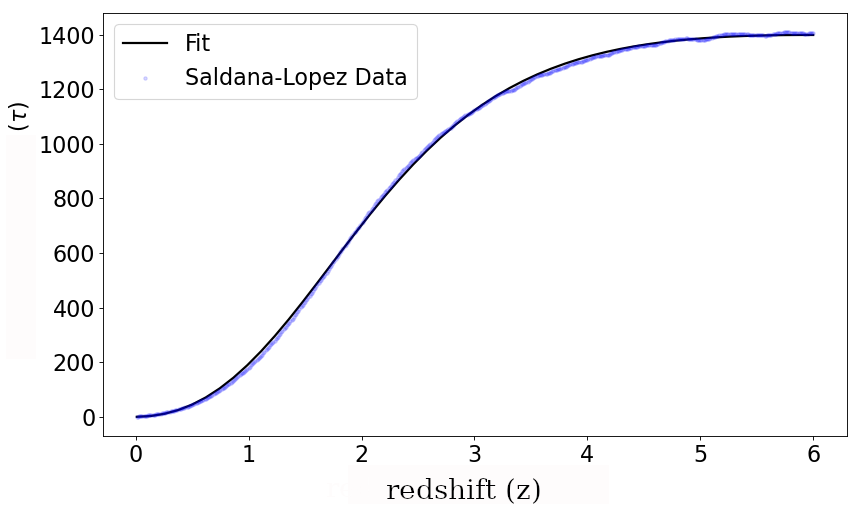}
\end{center}
\caption{\label{fig:optical}
Optical depth $\tau$ ~\cite{Saldana-Lopez:2020qzx} as function of cosmological redshift $z$ of its source for a photon of measured energy 10 TeV. 
A simple rational function fits the data rather well.}
\end{figure}

TeV photons in consequence are very strongly absorbed in the tenuous intergalactic medium (see Figure~\ref{fig:optical}) and stem from our cosmic neighbourhood only.
This sizeable absorption means in particular that Einstein Telescope detection of gravitational waves from deep cosmic distances at $z>1$ will not be accompanied by $\gamma$-messenger signals above 50-100 GeV.

\begin{figure} 
\begin{center}
\includegraphics[width=0.7\textwidth]{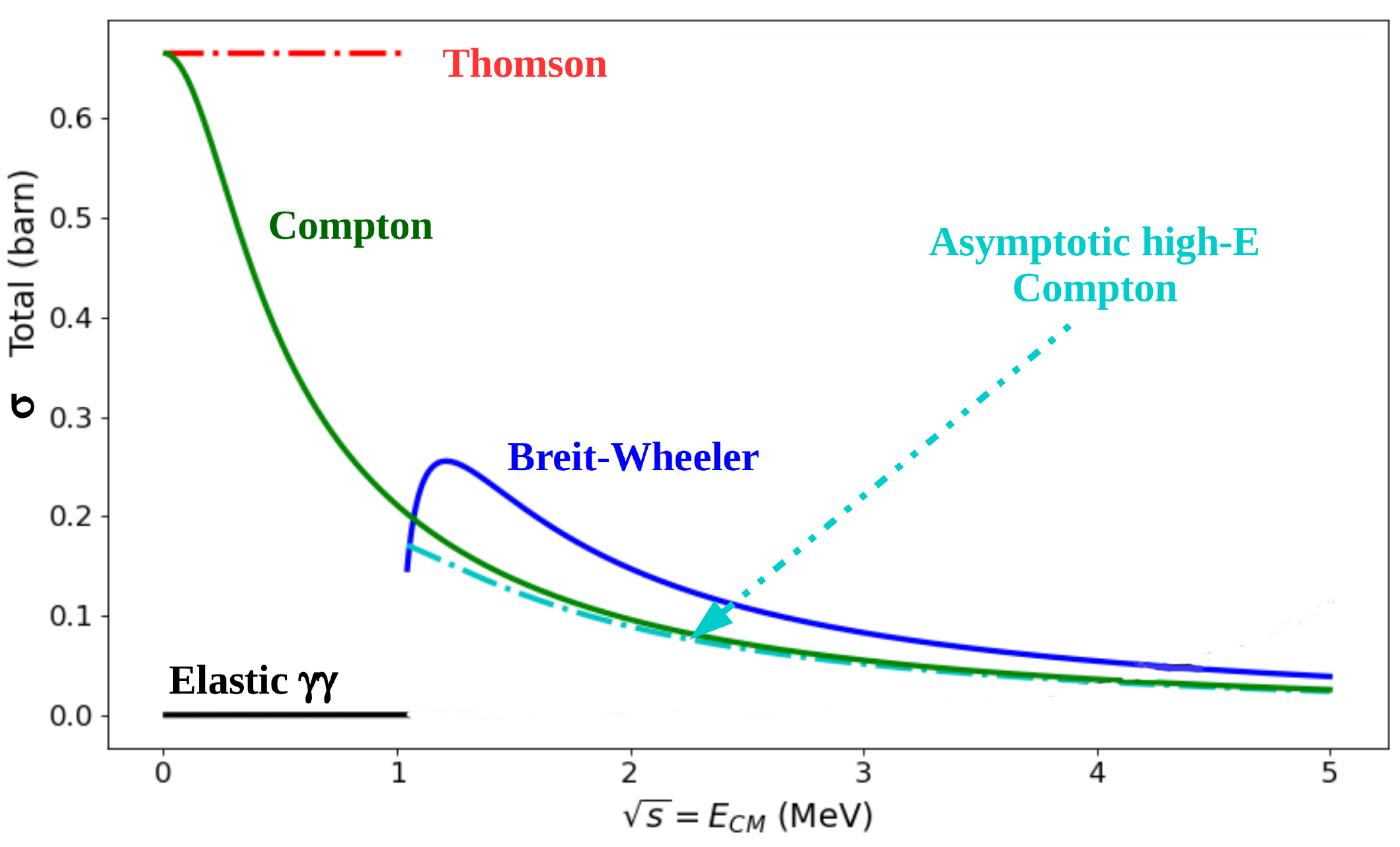}
\end{center}
\caption{\label{fig:cross} Cross-sections of relevant electrodynamics processes for the absorption of $\gamma$-ray photons
crossing intergalactic space.  }
\end{figure}

This absorption is due to quantum electrodynamics processes depicted in Figure~\ref{fig:cross}.

At high energies the Breit-Wheeler process $\gamma\gamma \to e^-e^+$ has the largest cross section, and because of the abundance of Extragalactic Background Light (EBL) photons to collide with, is the reason of the strong absorption in Figure~\ref{fig:optical}. 

At nuclear MeV energies, near and below the center of mass $e^-e^+$ threshold,  Compton scattering is dominant.  
The figure allows the appreciation of the decreasing cross sections $\sigma\propto 1/s$ as well as the energy-independent Thomson cross-section as the low energy limit. The electron density is identified with the average cosmological baryon density  (note that Thomson scattering on the proton is suppressed by $m_e^2/m_p^2$, so only electrons are considered).

It is thus remarkable that the noticeable absorption  of Figure~\ref{fig:optical} is not accompanied by a refractive delay of cosmic photons respect to 
the GW reference. Although tenuous, one could conceive that the intergalactic photon and electron medium might affect the real part $n_R$ of the effective refraction index $n$ of $\gamma$ rays, since the imaginary part $n_I$ must be sizeable and both real and imaginary parts are related by Kramers-Kronig dispersion relations. 

This work is therefore dedicated to clarifying the situation. We explicitly show a complete calculation of the dispersion relations and all the attending kinematic and dynamic factors that correct it. Our finding is that these additional factors include $E_\gamma/H_0$ (the ratio of the photon energy and the Hubble constant) that is of a huge magnitude in the cosmological context, and therefore explains the apparent discrepancy of small refraction and large absorption with the naive expectations from a dispersive analysis.

The calculation is presented in the rest of the terse article. 
Sections~\ref{sec:delay} and~\ref{sec:KK} are largely dedicated to high-energy photons in the TeV regime,
Section~\ref{sec:Compton} handles the low-energy regime where Compton scattering is dominant, and 
Section~\ref{sec:factors} dwells into the reason why absorption is so much larger than refraction. 
Finally, conclusions and a short comment on other scattering processes that turn out not to be so relevant are presented.

\section{Time delay, refraction index and dielectric constant} \label{sec:delay}

To center ideas, let us follow in this and the next two sections 
the chain of reasoning that leads to the computation of the time delay
for a high-energy photon, for which the pair-production process is the dominant scattering.
When a $\gamma$ photon from a cosmic source collides with a photon of the EBL~\cite{Cooray:2016jrk}, the center of mass energy
 is given by 
\begin{equation}
E_{\rm CM}=\sqrt{s}=\sqrt{2\epsilon_{\rm EBL}\epsilon_{\gamma}(1-\cos\varphi)} \quad .
\end{equation}
There, $\epsilon_{\rm EBL}$ represents the energy of the background photon, whose distribution depends on $z$
as given by the cosmological evolution (the photon energy redshifts with the Friedmann expansion parameter $a$, and their number density diminishes with $a^3$; but stellar emission replenishes that density, so we adopt the experimental determination).
Likewise, $\epsilon_\gamma = E_\gamma \times (1+z)$ is the $\gamma$-ray energy at the time (given by $z$) at 
which the collision would have taken place, and $\varphi$ is the collision angle between both incident particles.

Because the cross-section depends on this energy, as per Figure~\ref{fig:cross},
the time delay must be a function of the photon energy; it will also depend on the redshift through the density of photons 
$n_{\gamma}$. We obtain it as an integral along the line of sight: the total travel time would follow from integrating $dl/c$, and the delay from correcting $c$ by  $\frac{1}{n_R-1}$. Changing variables from distance $l$ to redshift $z$ yields
\begin{equation}
\Delta t\left(E_\gamma,z_0\right)=\int_0^{z_0} \frac{1}{c}\left|
 \frac{d \ell}{d z}\right| \left(n_R(\epsilon_{\gamma},z)-1\right)\ dz\ .
\end{equation}

As advanced in Section~\ref{sec:intro}, we will examine the Kramers-Kronig relations, that are more readily
written in terms of the dielectric constant $\epsilon$, so we square the complex refraction index,
\begin{equation}
\frac{\epsilon(\omega)}{\epsilon_0}\ =\ n^2\ =\ n_R^2-n_I^2-i2n_Rn_I \quad .
\end{equation}

Therefore, the real and imaginary parts of that ``dielectric'' response function (actually including 
the conductance), once linearized in the small $n_R-1:=\Delta n_R$, $n_I$, are
\begin{eqnarray*}
    \left\lbrace\begin{array}{c}
    \text{Re}[\epsilon(\omega)/\epsilon_0]=n_R^2-n_I^2\approx 1+2\Delta n_R \ ,\\
    \\
    \text{Im}[\epsilon(\omega)/\epsilon_0]=-2n_Rn_I\approx -2n_I \ .
    \end{array}\right.
\end{eqnarray*}

\section{Medium response: Kramers-Kronig relations} \label{sec:KK}

Dispersion relations are based on analyticity of response functions in the energy representation, 
that in turn follow from the causality of their Fourier-transformed time representation.
In the case of the electric response of a medium, the response is proportional to the applied field, with
proportionality factor $\epsilon$ that will depend on energy,
\begin{equation}
\vec{D}(\vec{x},\omega)=\epsilon (\omega) \ \vec{E}(\vec{x},\omega)\ .
\end{equation}

The Fourier transform is then
\begin{equation} \label{eq:DofE}
    \vec{D}(\vec{x},t)=\epsilon_0\left\lbrace \vec{E}(\vec{x},t) +  \int_{-\infty}^{\infty} dT \ G(T)  \  \vec{E}(\vec{x},t-T) \right\rbrace\ .
\end{equation}

In that expression, we have introduced the response function $G(T)$ that links the present displacement $\vec{D}$ to the past applied electric field $\vec{E}$ (the consequence follows the cause). This means that 
\begin{eqnarray}
   G(T)&=&  \frac{1}{2 \pi}\int_{-\infty}^{\infty} d\omega  [\epsilon (\omega)/\epsilon_0 - 1]  \  e^{-i\omega T}
\\  \epsilon (\omega)/\epsilon_0 - 1
  &=& \int_{-\infty}^{\infty} dT G(T) e^{iT\omega}\ .
\end{eqnarray}
If the frequency is extended to be a complex-space variable $\omega=\omega_R+i\omega_I$,  
$\epsilon(\omega)$ is upper-plane analytic for $\omega_I>0$; this is because $G(T)=0$ for $T<0$, so that the lower integration limit can be set to 0, and therefore the exponential contains a convergence factor $e^{-T |\omega_I|}$.

\begin{figure}
\begin{center}
\includegraphics[width=0.46\textwidth]{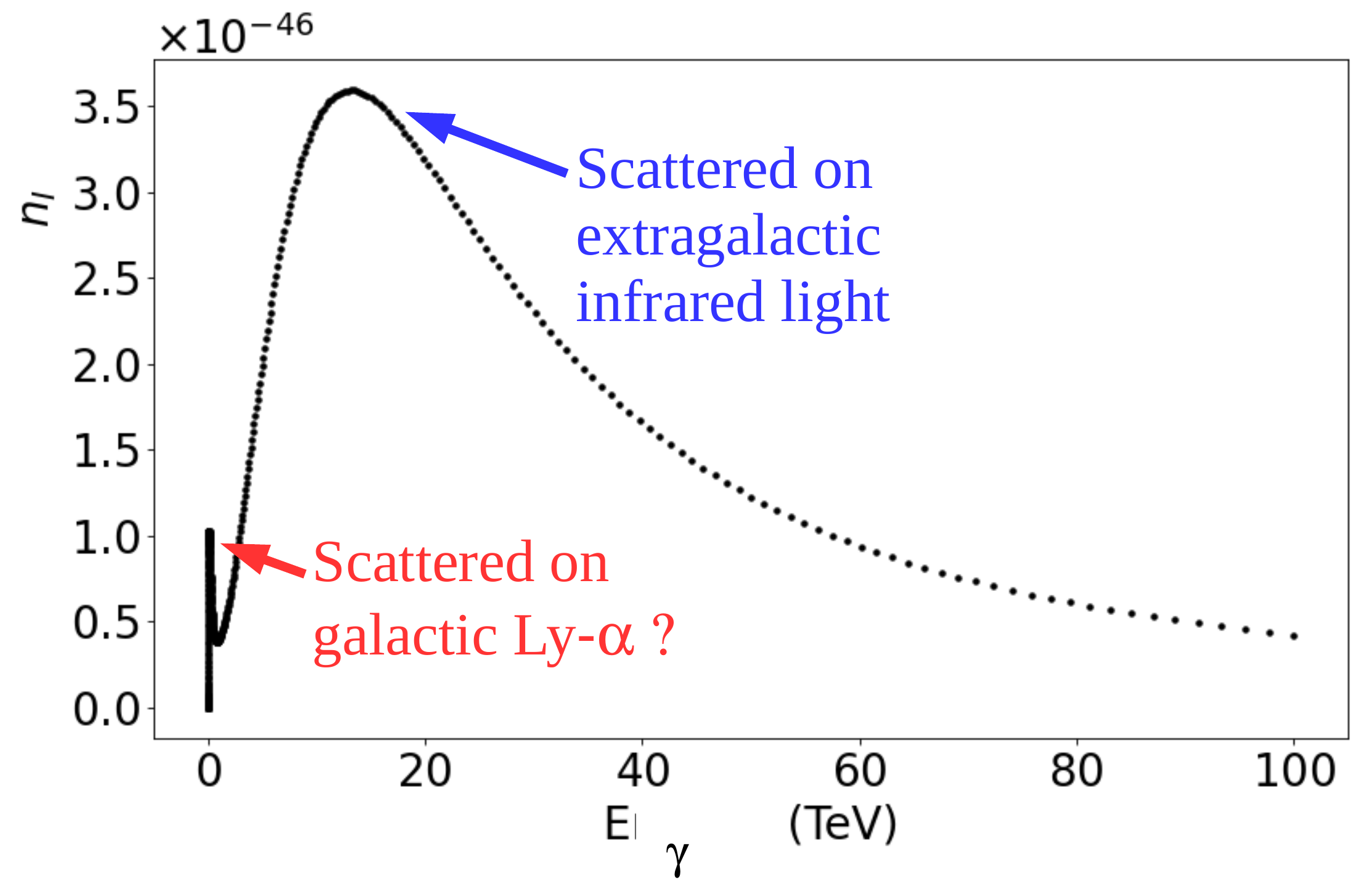} \raisebox{2cm}{ $\implies$} 
\includegraphics[width=0.46\textwidth]{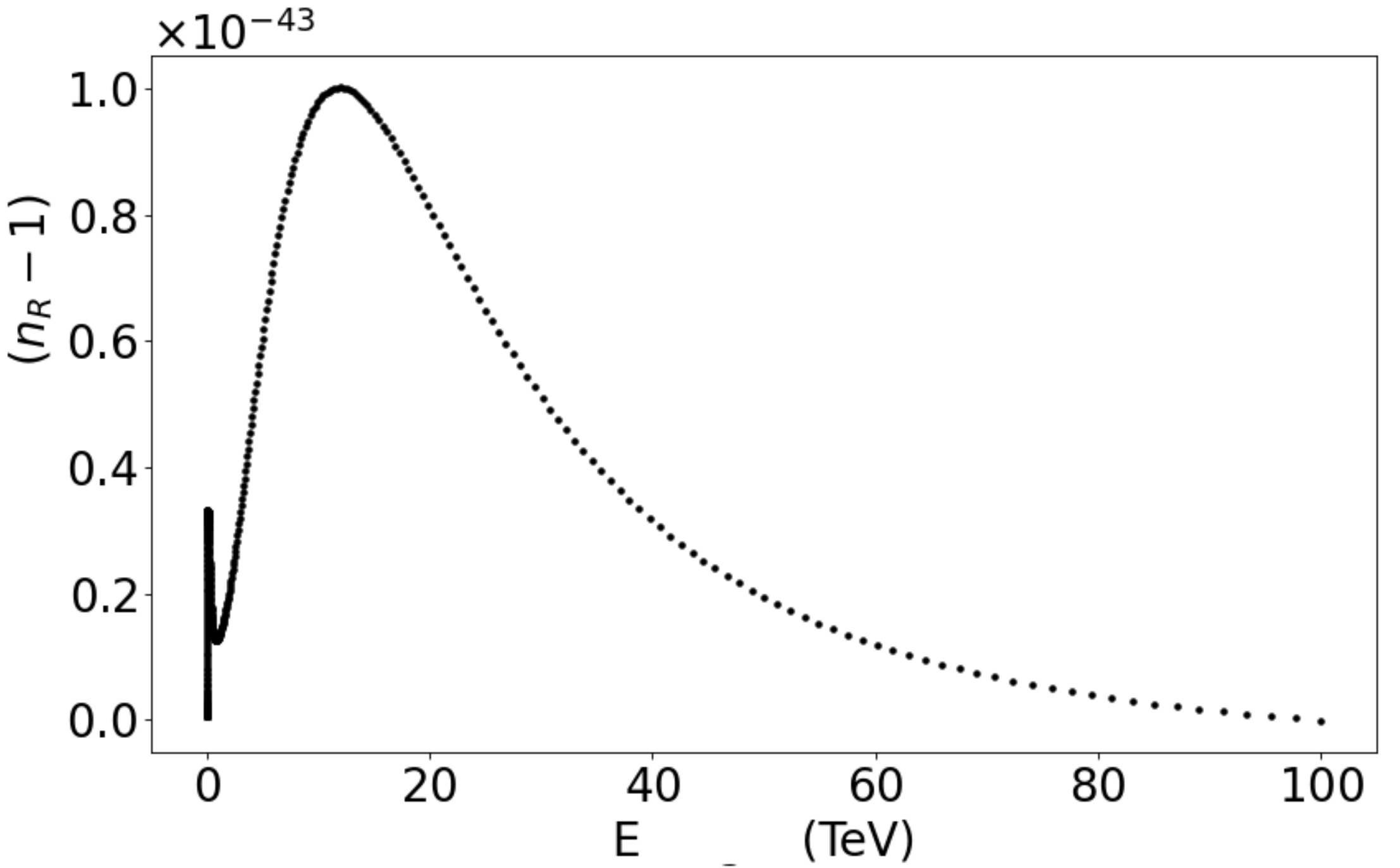}
\end{center}
\caption{\label{fig:n}{\bf Left:} imaginary part of the refraction index at $z=2$ obtained from the optical depth. {\bf Right:} real part of the refraction index as calculated with a Kramers-Kronig relation integrated from the left plot.
The broad peak in $n_I$, inherited by $n_R$, is maximum at an energy consistent with the Breit-Wheeler process on EBL photons that are today in the infrared or optical bands, relatively abundant because of star light that has escaped galaxies and travels intergalactic space. The narrow, weaker peak to its left (lower $\gamma$ energy in this plot requires higher $\epsilon_{\rm EBL}$) is likely adscribed to scattering off galactic Ly-$\alpha$ photons. The scales of both $n_R$ and $n_I$ are infinitesimally (but commensurably) small. }
\end{figure}

This analyticity allows the application of Cauchy's theorem to the upper-half plane in $\omega = \epsilon_\gamma/\hbar$; 
because of the convergent upper half-circle, what remains is an integral over the real axis, that expresses the dispersion relation. The resulting pair of Kramers-Kronig relations then read
\begin{eqnarray}
    \left\lbrace\begin{array}{l}
    \text{ Re}[\epsilon(\omega)/\epsilon_0]-1=   \frac{2}{\pi}\ \ {\rm PV} \int_{0}^{\infty} \frac{\omega' \text{ Im}[\epsilon (\omega')/\epsilon_0]}{\omega'^2-\omega^2} d\omega' \\
    \\ \\
    \text{ Im}[\epsilon(\omega)/\epsilon_0]= \ \ -\frac{2 \omega}{\pi}\ \ {\rm PV}  \int_{0}^{\infty} \frac{{\text Re}[\epsilon (\omega')/\epsilon_0] - 1}{\omega'^2-\omega^2} d\omega' 
    \end{array}\right. \ .
\end{eqnarray}
Focusing on the first one, the kernel has a $1/\omega'$ net factor for large $\omega'$,
so that integration convergence will succeed for $\epsilon(\omega')$ functions with a negative power-law exponent. This will be the case for the imaginary part of $n$ reported in the left panel of Figure~\ref{fig:n}.

In brief, the computation chain that allows us to predict the delay of a $\gamma$ photon proceeds from the optical
depth extracted from $\gamma$ absorption, to the real part of the refraction index that eventually reduces 
the speed of light, to, finally, the sought delay:
\begin{equation}
\tau \to n_I \xrightarrow{\rm Kramers-Kronig} n_R \to v =\frac{c}{n_R} < c \xrightarrow{\int dl } \Delta t\ .
\end{equation}

As already advanced, we need to change the variable of the integral along the line of sight, from the photon path-length to the measurable redshift,     
\begin{equation}
\int_{0}^{l_{z_0}} dl= \int_{0}^{z_0} dz \ \left| \frac{dt}{dz}\right| 
\end{equation}   
where the Jacobian is
\begin{equation}
\left| \frac{dt}{dz}\right|^{-1}=H_0 (1+z)^2\sqrt{\Omega_M(1+z)+\Omega_\Lambda (1+z)^{-2}}\ .
\end{equation}

Feeding the optical depth provided by high-energy experimentalists at our institute~\cite{Saldana-Lopez:2020qzx,servidorUCM}
we can numerically evaluate the refraction index, that turns out to be tiny. We quote it, for example, for scattering that would have taken place at $z=2$,  in Figure~\ref{fig:n}.
As for the shape of the curve, the main feature is a central peak due to EBL photons in the range 0.1-10 eV (with larger number towards the low-end of the interval as discussed by the Fermi-LAT collaboration~\cite{Fermi-LAT:2018lqt}) as well as a narrow peak due to scattering against Ly-$\alpha$ photons.

\begin{figure}
\begin{center}
\includegraphics[width=0.47\textwidth]{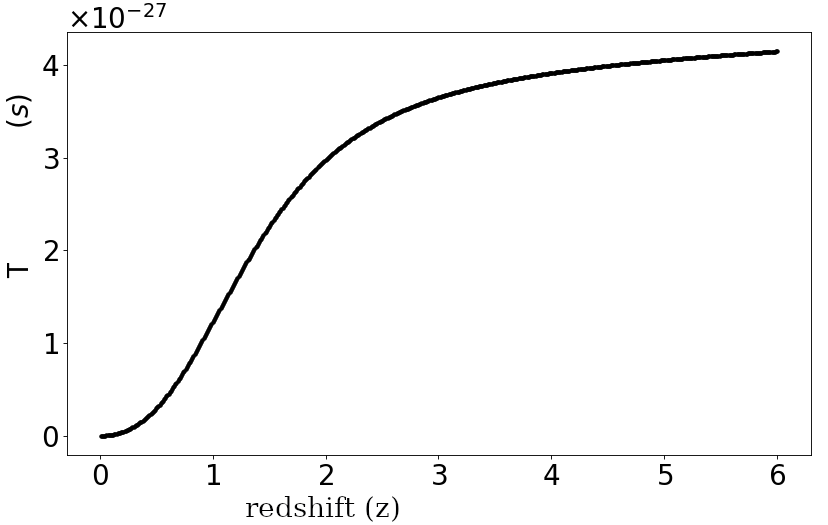}
\ \ \ \ 
\includegraphics[width=0.47\textwidth]{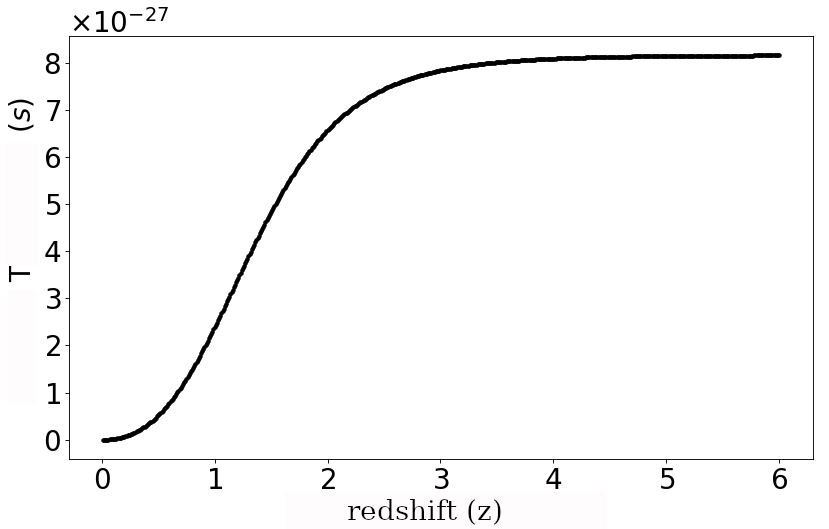}
\end{center}
\caption{\label{fig:delay} The delay times of $\gamma$-ray photons through the Extragalactic Background Light
are completely negligible; we plot them for a 100 GeV photon (left) and a 10 TeV one (right), as function of the red shift of their source. }
\end{figure}

With the refraction index at hand, we can obtain the time delay by numeric integration over $z$,
and we plot it in Figure~\ref{fig:delay} for detected 100 GeV and 10 TeV photons as a function of the source's $z$. 
The numerical values are completely negligible in all circumstances: accumulated delays are of the size of characteristic  subnuclear time scales, in spite of the cosmic travel lengths.

\section{Low-energy: Compton scattering $\gamma e\to \gamma e$} \label{sec:Compton}
Let us now turn to the calculation of the refraction for lowest energy $E\sim 0.1-1$ MeV photons, 
that might  possibly be scattered  on the way to Earth.

For this, the optical depth $\tau$ has not directly been measured, to our knowledge. Therefore, we turn to theory. 
Below the pair-production threshold, the dominant process would be Compton scattering, whose cross-section in terms of the energy at the needed $z$ where the scattering took place  is known, to Leading Order (LO), as a function of $x:=\frac{2\epsilon_\gamma(z)}{m_e}$ given by 
\begin{eqnarray}
 \sigma_{\rm LO}=\frac{\pi \alpha^2}{m_e^2(x+1)} & \phantom{+}& \left[ \frac{x^3+18x^2+32x+16}{x^2(x+1)} \right.\\ 
\nonumber &+& \left. 
 \frac{(2x^3-6x^2-24x-16)}{x^3} \ln(x+1) \right]\ .
\end{eqnarray}

Convenient limits to check the computer codes are the high-energy asymptotic behavior obtained by setting $m_e^2\ll s$ keeping in mind that
\begin{equation}
\frac{d\sigma}{d \rm{cos}\theta} \approx \frac{2\pi \alpha^2}{2m_e^2+s(1+\rm{cos}\theta)} 
\end{equation}
so that 
\begin{equation}\label{Comptonhigh}
\sigma_{\rm HE}=\frac{2\pi\alpha^2}{s}\ln\left(\frac{s}{m^2}\right)
\end{equation}
and the low-energy (Thomson) limit,
\begin{equation}\label{Comptonlow}
 \frac{d\sigma}{d\rm{cos}\theta}=\frac{\pi \alpha^2}{m_e^2}(1+\rm{cos}^2\theta);\ \ \  \ \ \ 
\sigma_{\rm Th }=\frac{8\pi\alpha^2}{3m_e^2}\ .
\end{equation}

While proceeding from theory we could try to directly calculate $n_R$, we prefer to pass by the imaginary part and use the same reasoning as for the high-energy photons to keep the unity of the discussion and be able to compare all steps. 
Therefore, we proceed to computing $n_I$ from the  optical depth, obtained from standard kinetic theory but substituting 
again the path length $l$ by the redshift $z$, which introduces the cosmological model (we take the matter content to be 
$\Omega_\Lambda=0.7$, $\Omega_M=0.3$ to saturate the cosmic sum rule, as radiation has been negligible for $z<10$), obtaining
\begin{equation} \label{opticalCompton}
\tau_{\rm{Compton}}=\frac{n_{\rm{barion},0}}{H_0}\int_{0}^{z_{\rm{max}}}dz' \ (1+z') \ \frac{\sigma_{\rm LO}(\epsilon_\gamma,z')}{\sqrt{(1+z')\Omega_M+(1+z')^{-2} \Omega_\Lambda}} \ .
\end{equation}

The needed electron density is obtained from charge neutrality and the known baryon density,
$N_e/V = n_{\rm{barion},0}= 0.25/m^3$, today, and scaled backwards with the volume $a^3=1/(1+z)^3$ factor as necessary.
The integral is easily computed by a quadrature rule, that has been checked analytically by simplifying the cosmology with only one matter content and using the limits of the Compton cross section in Eq.~(\ref{Comptonhigh}) and~(\ref{Comptonlow}).

\begin{figure}
   \centering
    \includegraphics[width=0.52\textwidth]{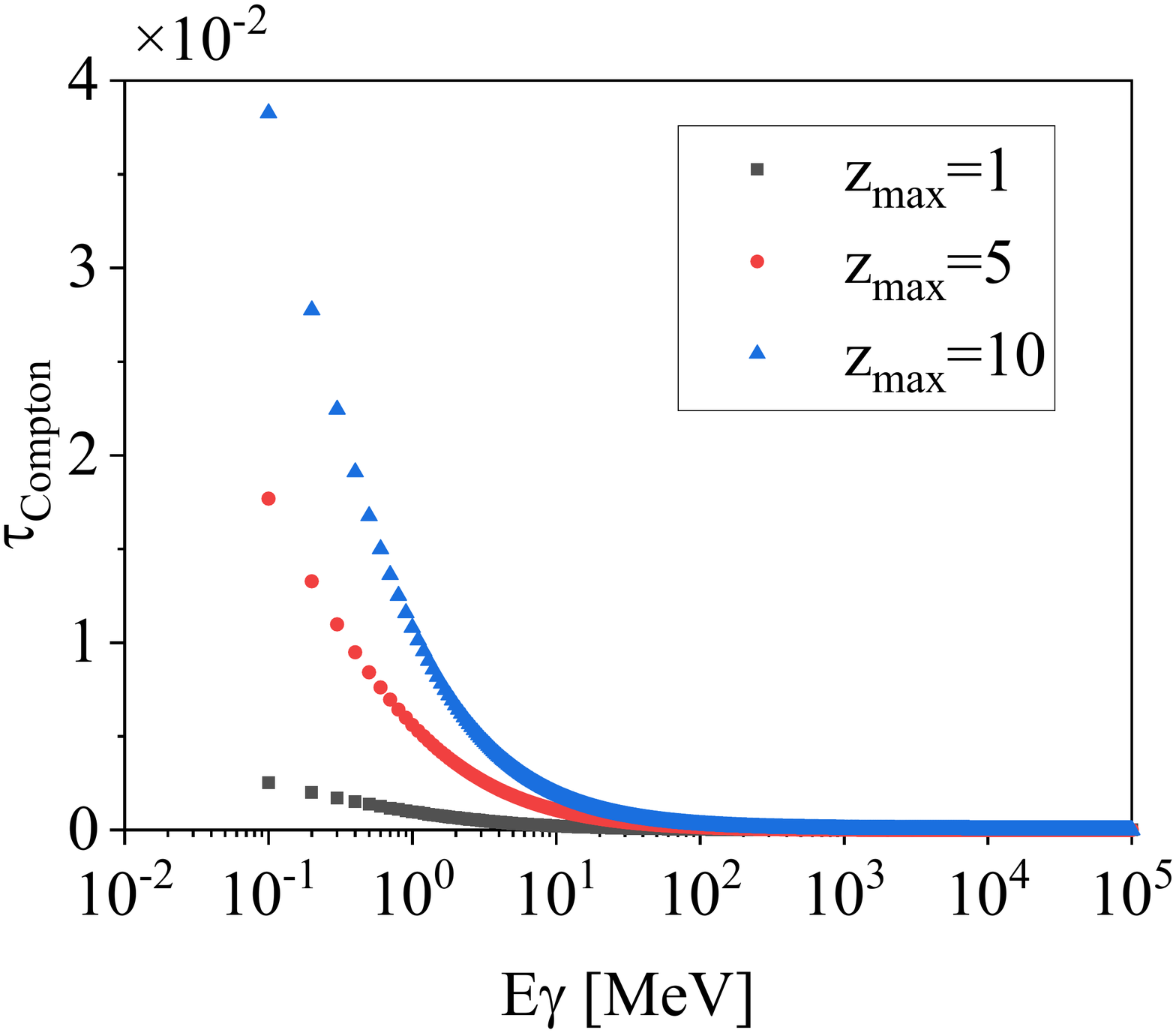}
    \caption{Compton optical depth $\tau_{\rm{Compton}}$ as function of photon energy $E_\gamma$, for three redshift values, $z_{\rm{max}}=1, \ 5, \ 10$.}
    \label{tauCompton}
\end{figure}

To obtain the imaginary part of the refraction index we note its relation with the optical depth, that is then plot in Figure~\ref{tauCompton},
\begin{equation}
d\tau=n_I k dr= n_I \left(\frac{2\pi c}{\lambda_{em}}\right) \left| \frac{dr}{dz}\right| dz = n_I \left(\frac{2\pi E_\gamma}{h} (1+z)\right) c \left| \frac{dt}{dz}\right| dz\ .
\end{equation}

We can then basically read-off $n_I$ from a derivative of Eq.~(\ref{opticalCompton}), resulting in
\begin{equation}
 n_I=\frac{n_{\rm{barion},0}}{E_\gamma}(1+z)^2\sigma_{LO}(E_\gamma,z)\ .
\end{equation}

The Kramers-Kronig relation in turn gives us $\Delta n_R =n_R-1$ as 
\begin{equation} \label{nRCompton}
 \Delta n_R = -\frac{2}{\pi} n_{\rm{baryon},0} (1+z)^2\ \  \rm{PV} \int_0^\infty \frac{\sigma_{LO}(E_\gamma',z)}{E_\gamma'^2-E_\gamma^2} dE_\gamma' 
\end{equation}
where the baryon density  $n_{\rm{baryon},0} =1,92 \ 10^{-39} \ [\rm{MeV}^3]$  is a tiny number when expressed in MeV$^3$ 
(as appropriate for Eq.~(\ref{nRCompton}) where the photon energy and cross-section are typically of order MeV and 
MeV$^{-2}$). The emptiness reflected in this low density is the ultimate reason for the smallness of both the imaginary and real parts of the refraction index.

From Eq.~(\ref{nRCompton}), the time delay can again be obtained for these low-energy $\gamma$ photons as
\begin{equation}
\Delta t =H_0^{-1} \int_0^{z} \Delta n_R(E_\gamma,z') \frac{dz'}{(1+z')^2\sqrt{\Omega_M(1+z')+\Omega_\Lambda (1+z')^{-2}}}\ .
\end{equation}

\begin{figure}
\begin{center}
\includegraphics[width=0.45\textwidth]{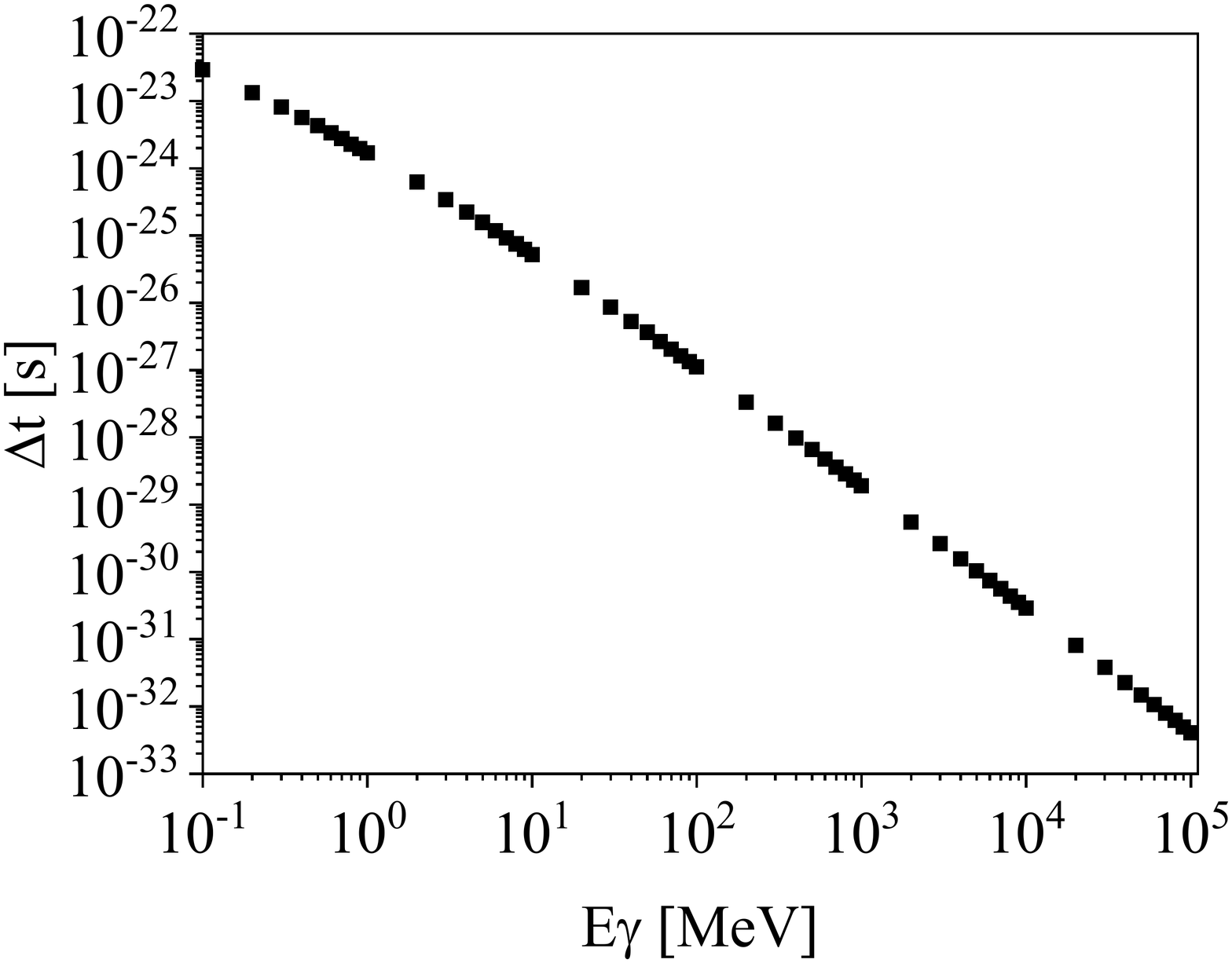}\ \ 
\includegraphics[width=0.45\textwidth]{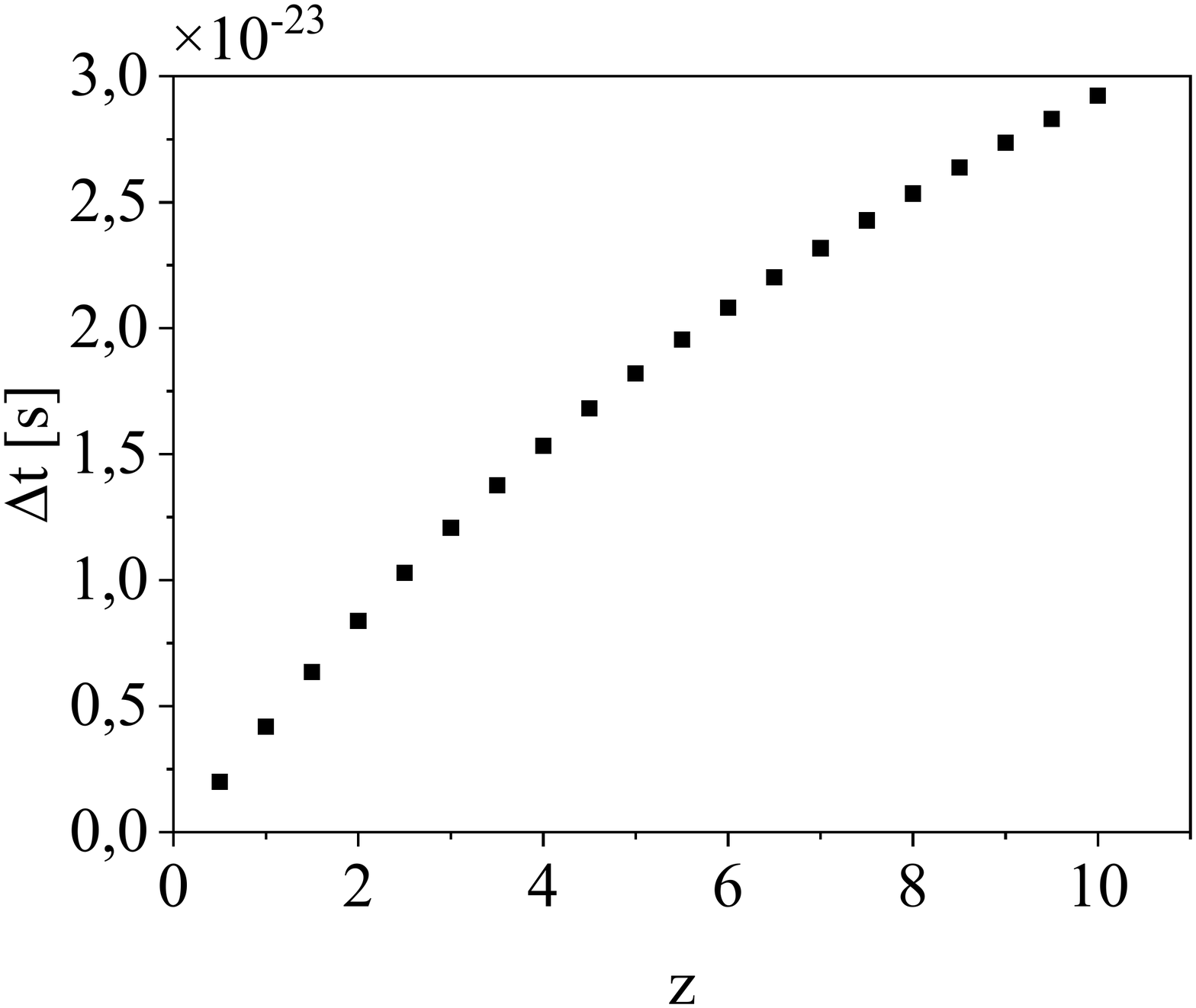}
\end{center}
\caption{Time delay of low-energy $\gamma$-rays. {\bf Left:} as a function of energy for a source emitting at fixed
$z=10$. {\bf Right:} as a function of $z$ for a fixed energy  $E=0.1$ MeV. As for high-energy photons, the delay is unmeasurably small and can be neglected for all applications. \label{fig:delayCompton}}
\end{figure}

Once more, we find that the time delay is numerically tiny, as exemplified in Figure~\ref{fig:delayCompton}.

\section{Absorption without refraction?} \label{sec:factors}

So far we have reasonably established that, because of the low target density given the typical energy and cross-sections, 
both $n_R$ and $n_I$ are very small. The finding is consistent, as we have seen, with negligible refraction/delay of $\gamma$ rays. 

This is in contrast with the nascent field of nuclear photonics in which, since the groundbreaking experiments at the Laue-Langevin institute~\cite{Habs} that established $\gamma$-ray refraction with index $n_R \simeq 1+ 10^{-9}$  (bending the beam by a millionth of a degree),
such refraction is ordinarily considered: the difference resides, of course, in the much larger density of the metals used to bend the laboratory beams. 

That would end the investigation were it not because of the sizeable absorption that we have seen for high-energy photons in Figure~\ref{fig:optical} and the smaller but not completely negligible absorption that follows for low-energy photons, see Figure~\ref{tauCompton}.

To understand the difference between this measurable absorption and the hopelessly unmeasurable delay we will next
compare the optical depth with the fractional time delay (both dimensionless quantities), obtaining
\begin{eqnarray}
    \tau &=& E_\gamma H_0^{-1}  \int_0^{z} n_I(E_\gamma,z') \frac{dz'}{(1+z')\sqrt{\Omega_M(1+z')+\Omega_\Lambda (1+z')^{-2}}} \\ 
& \phantom{=} & \\ & \phantom{=} & \\
    \frac{\Delta t}{t}&=&\frac{H_0^{-1}\int_0^{z} \Delta n_R(E_\gamma,z') \frac{dz'}{(1+z')^2\sqrt{\Omega_M(1+z')+\Omega_\Lambda (1+z')^{-2}}}}{H_0^{-1}\int_0^{z}\frac{dz'}{(1+z')^2\sqrt{\Omega_M(1+z')+\Omega_\Lambda (1+z')^{-2}}}} \ .
\end{eqnarray}

We see the respective proportionality of $\tau$ to $n_I$ and of $\Delta t/t$ to $n_R-1 = \Delta n_R$,  and how the optical depth is enhanced by a factor of the photon energy and the Hubble constant $E_\gamma H_0^{-1}$.

The  appearance of the energy is easy to understand, being typical of  the Bouguer-Beer-Lambert law in atomic spectroscopy.
Because the absorption coefficient of a plane wave is $\gamma = 2\omega n_I/c$, and also the derivative of the optical depth respect to the physical length, we can write
\begin{equation}
\underbrace{n_I}_{\rm Im.\ refraction} = \frac{\hbar c}{2} \frac{1}{E_\gamma} \underbrace{\times \ \ \frac{\partial \tau}{\partial l}}_{\rm attenuation\ coeff.}\ .
\end{equation}

Solving for the imaginary part of the refraction index and performing the change of variables $l\to z$ then yields
the additional factor of the Hubble constant,
\begin{equation}
n_I = \frac{\hbar c}{2} \frac{1}{  E_\gamma}  \frac{\partial \tau}{\partial z} \frac{ H_0(1+z)}{c}\ .
\end{equation}

We may then compare the expressions for the optical depth and relative time delay,
\begin{eqnarray}
\underbrace{n_I}_{\rm small} \underbrace{ E_\gamma H_0^{-1}}_{\rm large} &\propto &\underbrace{\tau}_{O(10^{-3}-10^3)}
\\
\underbrace{n_R}_{\rm small} \phantom{ \underbrace{E_\gamma H_0^{-1}}_{\rm large}} &\propto& 
\underbrace{\Delta t/t}_{\rm small} \ .
\end{eqnarray}

We may numerically evaluate the ratio of the two quantities,
\begin{equation}
\frac{\Delta t/t}{\tau}\propto \frac{H_0}{E_\gamma} =  
\frac{[10^{-20}, \ 10^{-26}]{ \rm s}}{(4,54\pm0,13) \ 10^{17} { \rm s }}
\end{equation}
which shows that, indeed, the optical depth can be far larger than the time delay.

\section{Conclusions}

In  conclusion, we have shown in this work that $\gamma$-rays do indeed ``run on time'', that is, they do not 
accumulate any appreciable delay respect to gravitational waves during their propagation from an astrophysical source
such as a neutron star merger~\cite{Llanes-Estrada:2019wmz} or a blazar. 
Our input information is the standard cosmological model with nonrelativistic matter and $\Lambda$, understanding of electrodynamics cross-sections, dispersion relations, and the absorption data for $\gamma$-rays obtained in the last few years.

This lack of delay is in spite of the  sizeable $\gamma$ absorption, especially at highest energy, that leads to the formation of a horizon for 100 GeV+ $\gamma$s that are lost over cosmic distances and will not accompany GW pulses.

We have reported an explicit computation with the  Kramers-Kronig relations suggesting, as expected, that $n_R\sim n_I$ within a factor of a thousand.

Therefore, the stark contrast between refraction and absorption must be given by the additional factors external to the refraction index (that is, itself, very small due to the rarified intergalactic gas). We have identified these factors and clarified the large ratio between absorption and refraction.

Finally, we expect no delay between GWs and MeV-TeV $\gamma$s due to cosmological propagation through the entire ET field of view, whenever such photons can be detected.

\acknowledgments
Supported by grants  MICINN: PID2019-108655GB-I00, PID2019-106080GB-C21 (Spain); UCM research group 910309 and the IPARCOS institute.


\newpage
\appendix
\section{Other  scattering processes} 
In the first place, let us comment on photon-photon scattering below the threshold for electron-positron pair production, that we have neglected against Compton scattering on cosmic electrons in spite of the larger abundance of photons.

This is easy to understand, as the creation and absorption of virtual pairs, that imply the nonlinear corrections to the Maxwell equations, controlling $\gamma\gamma\to \gamma\gamma$, can be encoded in the 
Euler-Heisenberg Lagrangian
\begin{equation}
\mathcal{L}_{E-H}=\frac{\alpha^4}{90m_e^4} \left\{  \left(F_{\mu\nu}F^{\mu\nu}\right)^2+\frac{7}{4}\left(F_{\mu\nu}\tilde{F}^{\mu\nu}\right)^2\right\}
\end{equation}
that leads to the cross-section given by 
\begin{equation}
\sigma_{\gamma\gamma\rightarrow\gamma\gamma}=\frac{973(\hbar c)^2}{10125c^{16}}\frac{\alpha^4\sqrt{s}\ {}^6}{m_e^8} \quad .
\end{equation}

It is straightforward to see that this is suppressed respect to the characteristic Thomson cross-section for 
$\gamma e\to \gamma e$ by  a factor $\propto \alpha^2 E^6_{\rm CM}/m_e^6$ that is extremely small for $100$ keV photons.

Additionally, a comment is warranted on the propagation of gravitational waves. These can also be scattered by mass distributions.
In the quasiNewtonian approximation, the metric can be approximated by
\begin{equation}
    g_{00}=(1+2\Phi), \ \ g_{0i}=0, \ \ g_{ij}=-\delta_{ij}(1-2\Phi)\ .
\end{equation}
Then, one can compute the geodesics and interpret the trajectories of the GW ray vectors as bending by an
effective refractive index, known to be, in function of the energy density distribution $\rho$ of the background by the following expression, that yields a very small time delay,
\begin{eqnarray}
n=1+\frac{2\pi G \rho}{\omega^2}\to \Delta n \to \nonumber \\  
\Delta t = \frac{\Delta l}{c}(n-1) = \frac{10^{-18} ({\rm Hz})}{\omega^2}  \ .
\end{eqnarray}

This expression shows the same dependence on frequency than the Drude-Lorentz model for $\gamma e$ scattering in a material with a free electron cloud, with the obvious substitution of the gravitational force constant and energy density  by the
electric force constant and charge density.
At any rate, the result is extremely small for all practical cases: GW scattering can be safely ignored through intergalactic space too. The exception is lensing by macroscopic objects such as whole galaxies or cumuli; but in that case, because of the equivalence principle, both GWs and $\gamma$-rays follow the same geodesic around those objects and no relative delay between them is expected. In consequence, such delay would have been due to the electromagnetic scattering of the $\gamma$-rays, and we have shown that this is completely negligible.

\newpage


\end{document}